\pdfoutput=1
\documentclass{article}

\usepackage{amsthm}

\usepackage{amsmath, amssymb}           % Math and symbols
\usepackage{times}    
\usepackage{graphicx}                   % For images
\usepackage{booktabs}                   % Better tables
\usepackage{csquotes}                   % For \enquote
\usepackage{url}                        % URL support
\usepackage{setspace}
\usepackage{ragged2e} 
\justifying
\usepackage[hidelinks]{hyperref}       % URLs without colored boxes
\usepackage{epigraph}                  % Epigraphs
\usepackage{authblk}                   % Author affiliations
\usepackage{microtype}                 % Better spacing
\usepackage[margin=1in]{geometry}    % Wider margins
\usepackage{footmisc}

\usepackage[square,authoryear]{natbib}
\title{\Large\textbf{From Hallucination to Reliability: Generative Modeling and the Structure of Scientific Inference}}

\author{Charles Rathkopf\\ \normalsize Forschungszentrum J\"ulich}

\date{August 2026}

\begin{document}
\maketitle

\epigraph{
    {``But in the practice of science, knowledge is an affair of \textit{making} sure, not of grasping antecedently given sureties.''} 
}{John Dewey\footnotemark}

\footnotetext{\cite[p. 154]{dewey1958Experience}. [Emphasis in original.]}

\newpage
\begin{abstract}
Generative AI is increasingly used in science, but is unavoidably prone to hallucination. I develop a reliabilist account of how generative AI nevertheless gives rise to new scientific knowledge. I analyze hallucinations as non-strategic misrepresentations of the target phenomenon, introduced by a model's generative activity, rather than inherited from training data. Through case studies of AlphaFold and SEEDS, I show how scientific workflows draw on pre-existing knowledge of target phenomena to filter or qualify hallucinatory outputs, thereby preventing their erroneous content from propagating into downstream inference. Finally, I show that workflows are units of epistemic evaluation in their own right.
\end{abstract}
\bigskip
\begin{flushleft}
\textbf{Word count:} 8,100 (main text, abstract, footnotes, and figure caption; section headers and references excluded) 
\end{flushleft}

\newpage

\section{Hallucination as a threat to reliability}

In recent years, generative AI has become deeply embedded in scientific practice. It is now used to synthesize data for climate models \citep{kadow2020Artificial}, to map phase transitions in novel materials \citep{arnold2024Mapping}, and to predict molecular interactions for drug discovery \citep{sidhom2022Deep}. Unlike classificatory AI models, generative AI models produce outputs that are highly detailed and informationally rich. That richness makes them epistemically valuable, but also leaves them susceptible to a new kind of error that has come to be known as \enquote{hallucination} \citep{ji2023Survey, sun2024AI}.
 
As a first pass, hallucinations\footnote{The term invites a misleading comparison to human perceptual experience. Generative AI models do not consciously perceive the world, let alone misperceive it. Nevertheless, the term is now entrenched in the literature, and I suspect that insisting on a replacement would be futile.} can be characterized as errors that are not merely inherited from the training data, but are, in some sense, produced by the model itself. This claim, sketchy though it is, suffices to show that not all errors count as hallucinations. For example, if the training data includes a mislabeled datapoint, and the model reproduces it, then the output will be an error, but not a hallucination. I develop this characterization below, but even this initial sketch helps explain why the phenomenon demands attention. It is natural to worry that any model prone to hallucination will mislead researchers. Indeed, the epistemic risks are serious. AlphaFold 3, among the most celebrated generative models in science, has been shown to produce detailed molecular structures where none exist \citep{abramson2024Accurate}. GANs used in medical imaging have introduced phantom anomalies such as a fracture-like line in an unbroken bone, or a lesion in healthy tissue \citep{shin2021Deep}. 

These are not just rounding errors. Undetected, hallucinations can lead researchers and clinicians toward serious mistakes in inference and decision-making.

Moreover, there is reason to think that hallucinations are \textit{inevitable} byproducts of the mechanisms of generative inference. The intuition behind this claim is that training such models involves a fundamental tradeoff between novelty and reliability \citep{sajjadi2018Assessing, sinha2023Mathematical, xu2024Hallucination}. A model constrained to strictly mirror its training data may be reliable but also incapable of generating novel insights. Allowing a model to extrapolate, by contrast, enables novelty but also invites fabrication.

Another aspect of the epistemic threat hallucinations pose is that they are sometimes difficult to detect \citep{ji2023Survey, bubeck2023Sparks}. This is not always the case. When the target phenomenon is familiar, hallucinations can be easy to spot. For example, earlier versions of DALL-E and Stable Diffusion often generated images of human hands with six fingers \citep{wang2024MixtureofHandExperts}. But scientific AI operates at the frontiers of human knowledge, where familiarity is unavailable and errors are likely to go undetected. And the longer they remain undetected, the more they threaten to derail decisions based on model outputs.

Drawing these observations together, it seems that at least some hallucinations are substantive, inevitable, and difficult to detect. Compounding this problem is the fact that deep learning models are epistemically opaque \citep{humphreys2009Philosophical, creel2020Transparency, beisbart2026Which}. In traditional closed-form models, evidence of reliability is often grounded in knowledge of how parameters relate to properties of the target system. An error in the output can often be traced to a particular parameter and corrected there. In DNNs, by contrast, it is unclear whether individual parameters represent anything at all. As a result, we cannot assign blame for an error to any particular parameter. 

The problem of hallucination and the problem of opacity compound one another. Because hallucination is an inevitable byproduct of generative inference, no refinement of the architecture will yield a model that does not hallucinate. And because the model is opaque, the hallucinations it produces cannot be traced to any particular operation within the model that might be corrected. There is, therefore, no direct pathway from the discovery of an error to the construction of a more reliable model.

However, these models have demonstrated their ability to outperform traditional modeling approaches in all sorts of important predictive tasks. So the pressing question is not \textit{whether} generative AI models can be reliable, but \textit{how} the inferences we draw from them can be, given that hallucinations can be neither eliminated nor traced. The central aim of this paper is to make progress toward answering this question. 

Reliabilist epistemology \citep{goldman1979What, lyons2019Algorithm} can be read as suggesting a straightforward answer: \textit{just observe its track record.} Instead of explaining why a model succeeds, we can simply infer its reliability from past performance. The obvious concern with this approach, which \citet{duede2023Deep} calls \textit{brute inductivism}, is that it reduces scientific epistemology to an accounting exercise. High accuracy on benchmarks or alignment with historical data is no guarantee of future performance \citep{grote2024Reliability}. 

Nevertheless, I will argue that the most promising answer is to be found in the theory of \textit{computational reliabilism}~\citep{duran2018Grounds, duran2025Defense}, which says that a reliable computational process can confer epistemic warrant even if the process is opaque. The two case studies extend this idea by showing that, for generative AI, the relevant unit of epistemic evaluation is the error-mitigating workflow in which the model is embedded, rather than the generative model itself. The larger workflow is constructed on the basis of independently established knowledge of the target system, such that model outputs can be filtered or qualified before they are used in downstream scientific inference. 

I develop this account as follows. Section 2 provides two arguments for the inevitability of hallucination. Section 3 develops an analysis of hallucination according to which a hallucination is an output that (i) misrepresents its target phenomenon under the canonical interpretation of outputs of that kind, (ii) does so non-strategically, and (iii) owes its misrepresentational status to the model’s generative activity rather than to the training data or to inference-time inputs. The first condition contrasts with an alternative view, sometimes favored in the machine learning literature, according to which an output counts as hallucinatory when it deviates from the training distribution. This contrast is crucial to the view developed here because pre-existing scientific knowledge of the target system is precisely what enables researchers to build workflows that filter or qualify model outputs. Sections 4 and 5 provide two central case studies (AlphaFold and SEEDS) that demonstrate how generative AI models can support reliable inference when embedded in appropriate workflows. Section 6 addresses the objection that generative AI need not be reliable because it is in the business of discovery rather than the business of justification. Section 7 situates the account developed here in relation to the existing literature on computational reliabilism and argues that reliability-saving workflows can only be constructed in domains in which sufficiently mature scientific knowledge of the target system is already in place.

\section{On the inevitability of hallucination in generative AI}

\subsection{What is generative AI?}

The term \textit{generative AI} is sometimes taken to refer to any AI system that mimics the cultural products of human creativity. While many generative models do exactly that, mimicking human output is just one of many ways these architectures can be deployed. They are also used to produce numerical, physical, and scientific data of all kinds. Here is a definition that is sufficiently abstract to capture this broader scope:

\begin{quote}
A \emph{generative AI model} is a machine learning system trained to produce complex data structures by sampling from a distribution learned from the training data, while generalizing beyond the exact instances in that data.
\end{quote}

The word \enquote{complex} is carrying a lot of weight. The complexity of model outputs plays a central role in understanding both why hallucinations are inevitable and why they pose a distinctive epistemic threat in scientific applications. Here, \enquote{complexity} refers to high dimensionality: model outputs are structured, multi-component entities rather than scalar values or discrete labels. In many generative models, output dimensionality is proportional to that of the training examples. In some cases, such as \textit{SEEDS}, input and output have equal and fixed dimensionality by design, since both represent meteorological fields over a grid on the Earth's surface. In others, such as autoregressive language models, outputs may exceed the length or complexity of the inputs (e.g., \enquote{Write me an essay about the history of AI}). But even in those cases, outputs are bounded by factors like the maximum size of the context window.

In both kinds of case, the generative task involves producing plausible outputs in a high-dimensional space whose structure is incompletely determined by the training distribution.

Two contrasts help clarify what makes generative AI distinctive. The first is between generation and classification. A classification model learns to represent the conditional distribution \( P(Y \mid X) \) over discrete labels \( Y \). Generative models aim to learn a representation of the full distribution \( P(X) \), which enables them to produce novel samples that were not included in the training distribution \citep{kingma2013Autoencoding, goodfellow2014Generative, buckner2024Deep}. Second, unlike classical generative statistical models such as Poisson processes or Markov chains, which generate data from predefined parametric distributions \citep{grimmet1992Probability}, generative AI models learn latent representations that capture complex and often idiosyncratic statistical structure \citep{rezende2014Stochastic, yang2023Diffusion}.

\subsection{Inevitability arguments}

If hallucinations could be eliminated simply by retraining the generative model more carefully, or by adjusting its architecture, there would be no need to ascend to the workflow level of analysis. However, there are good reasons to think that these kinds of adjustments cannot eliminate hallucination. A substantial literature already argues for the inevitability of hallucination in large language models, drawing on no-free-lunch, incompleteness-style, and information-theoretic considerations \citep{xu2024Hallucination, banerjee2024LLMs, kalai2024Calibrated}. Those arguments are tailored for autoregressive AI models, where content is generated in small chunks, appended to what was generated in an earlier round, and then fed back into the model at each subsequent round. These models have played an important role in scientific applications, but recently, a different family of models has become even more important. These models, which include diffusion models, VAEs, and GANs, are trained to generate high-dimensional structures as wholes rather than one small unit at a time. The inevitability arguments for this class of models have not received much attention in the literature, so I will now adapt some of the existing ideas to diffusion models and related architectures.

The first argument is information-theoretic. Because these generative models have complex outputs, the size of the space of possible outputs is astronomical. No realistic training data distribution carries enough information to specify which portions of the space correspond to genuine possibilities. For example, there are \(10^{130}\) possible amino-acid sequences of length 100 \citep{dryden2008How}. Meteorological models are also designed to cover astronomical possibility spaces. Because a generative model is expected to generalize beyond the exact instances in its training data, it must generate outputs in regions that those data constrain only weakly. Some of the model's extensions into these underdetermined regions will therefore be misrepresentations.

A second argument is geometric. Generative models attempt to learn the geometry of high-dimensional data and generate new outputs by sampling from their learned distributions. But in high-dimensional settings, geometric properties become unintuitive. As \citet{arjovsky2017Wasserstein} note, real data typically lie on low-dimensional manifolds within a much larger ambient space. When generative models are trained on data that contain multiple such manifolds, interpolation can result in outputs that fall \textit{between} them. Then, when a model samples from these regions, hallucinations are generated.

Perhaps the most unnerving property of hallucinations is that they often have a kind of superficial plausibility. Hallucinations would matter less if they were obvious. At the scientific frontier, they often are not. Generative models often capture local dependencies more faithfully than global ones. Local structure is recurrent and statistically robust; long-range dependencies are subtler and harder to learn. As a result, generative models produce outputs that are locally plausible yet globally flawed. A weather forecast might look accurate, but contain violations of some conservation law, and a language model might deliver coherent sentences, but an incoherent argument.

These considerations motivate a general conclusion: any generative model that aims to produce complex, structured data will sometimes produce hallucinations. Moreover, contrary to what the recent success of AI scaling laws might suggest, even massive increases in the size of the training data will not make hallucinations of this kind go away.

\section{Rethinking hallucination}

So what is a hallucination, exactly? The most developed answer in the machine learning literature concerns diffusion models, and it will serve here as a foil. Diffusion models are, in any case, the most natural place to start because they have been responsible for many of the most widely discussed successes of generative AI in science, and both of the case studies below are built on them.

A diffusion model generates new data by progressively removing noise. At each stage, it takes a noisy data structure as input and produces a slightly less noisy version. To train the model, examples from a dataset are corrupted with varying degrees of Gaussian noise, and the model learns to reverse this process. The denoising process is guided by the \textit{score function}, the gradient of the distribution’s log-density, which the model learns to approximate. Crucially, neural networks tend to learn \textit{smooth} approximations of the score function even when the true score function contains sharp transitions.

A recent study by \citet{aithal2025Understanding} analyzes small diffusion models in order to understand why hallucinations occur. Their view is that the smoothness of the approximation of the score function induces interpolation across low-density regions of the data distribution, and this is what causes hallucination.\footnote{The score function of a distribution \( q(x) \) is defined as the gradient of its log-density: \( \nabla_x \log q(x) \). This function describes the changes in the probability density at each point \( x \). In real-world distributions, the function may contain sharp transitions, such as those separating distinct configurations of a hand. But neural networks approximate this function with a smooth one, which causes them to interpolate across the gaps between modes. Interpolation of that kind can then lead to hallucinatory images, such as an anatomically incoherent hand. For further explanation, see \citet{luo2022Understanding} or \citet{song2021scorebased}.}

To make this idea precise, they introduce a threshold-based definition of hallucination, according to which outputs that fall in sufficiently low-density regions of the data distribution are flagged as hallucinations.\footnote{

\begin{equation}
H_\epsilon(q) = \{x \mid q(x) \leq \epsilon\}
\end{equation}

Here, \( q(x) \) is the probability density of the real data distribution at output \( x \), and \( \epsilon \) is a small threshold.}

Aithal et al.\ also propose that hallucinations, so defined, can be discarded. They introduce a metric based on the variance of the model's prediction \( \hat{x}_0 \) over the sampling trajectory, treating unusually high variance as an indicator of hallucination. Samples whose trajectory variance exceeds a chosen threshold are discarded. According to their evaluation, this method removes 95\% of hallucinatory samples while preserving the vast majority of in-distribution outputs.

This data-centric solution is reasonable for synthetic datasets where the data-generating process is known (as it is in the experiments Aithal et al.\ report). But scientific modeling aims at discovering structure \textit{not} explicit in the training data. In that setting, a sample between known modes may signal \textit{either} modeling error \textit{or} discovery. For that reason, \citet{aithal2025Understanding}’s solution is too conservative: it risks eliminating precisely the outputs from which novel knowledge might emerge. If hallucination were simply a matter of deviation from training data, then some of the outputs that deserve to be regarded as discoveries would, \textit{ipso facto}, count as hallucinations, and then get filtered out.

The problem is not merely that Aithal et al.\ are too conservative. It is that conformity to training data is the wrong standard to pay attention to. Scientific inquiry should be designed to provide knowledge of the target phenomenon, and in order to ensure that our inferential practices help us to achieve that end, the danger we must guard against is model outputs that misrepresent the target phenomenon. 

The idea that inference is properly aimed at the phenomena behind the data, rather than at the data themselves, is a lesson that philosophers of science learned decades ago. Early logical empiricist projects, especially Carnap's \textit{Aufbau} \citep{carnap1928Aufbau}, treated science as the reconstruction of observational data. That view withered under criticism, but the underlying idea that theories earn legitimacy only by recovering or predicting patterns in data of some sort seems to have been widely accepted through the 1960s. But as \citet{bogen1988Saving} forcefully argued, this outlook fails to account for the constructed nature of data. This is not to deny that access to phenomena is mediated by data. Rather, it is a reminder that, because data sets are shaped by the contingencies of measurement techniques and experimental design, if you want to make good inferences about the phenomena, you need to attend to the process by which the data set was constructed \citep{woodward2011Data}. To put this thought in slogan form, the data are a means to an end, rather than an end in themselves.

Once we accept this orientation, we arrive at a different picture of how generative AI models ought to be assessed. A model's training data do not define the limits of its validity. What matters is whether its outputs support correct inferences about the target phenomenon. This idea is visualized in Figure 1. The relationship between a model's output and the training data is one of statistical resemblance; the relationship between the output and the target phenomenon is representational.\footnote{\citet{termine2024Machine}, \citet{kieval2026Representation}, and \citet{freiesleben2026Artificial} develop related positions, each in different ways, on how machine learning outputs relate to their targets: the statistical relation a model bears to its training data is not, on its own, a representational relation to the phenomenon the data were drawn from.}
\begin{figure}[ht]
\centering
\includegraphics[width=0.9\textwidth]{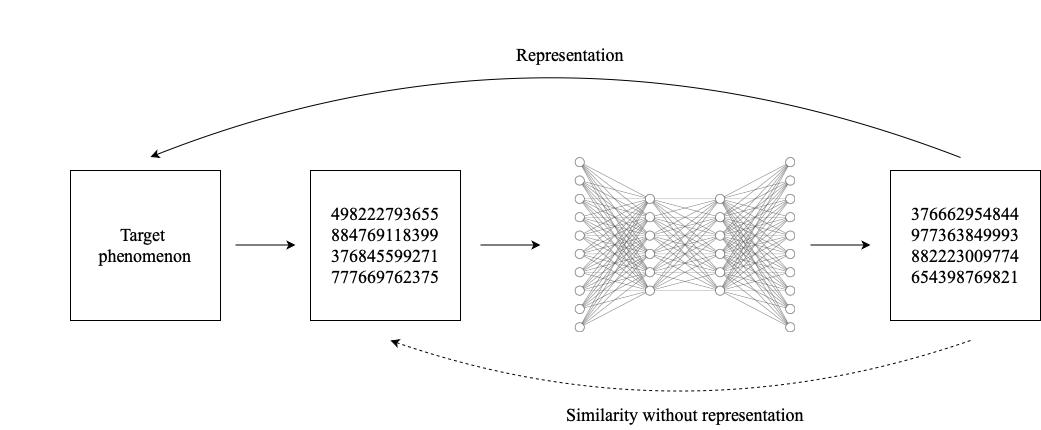}
\caption{
Diagram illustrating the relationship between a target phenomenon, a dataset constructed from observations of the target (second box), a generative deep neural network (DNN), and the DNN’s output. The DNN produces outputs that resemble samples from the training data but do not represent them. Whether an output functions as a representation depends on our inferential practices, and in scientific contexts, these practices are aimed at understanding the target phenomenon—not merely reconstructing the training data. The backward arrow (``similarity without representation'') indicates that while the model output may exhibit statistical similarity to training data, it is not used as a representation of the training data itself. Rightward arrows indicate causal rather than representational relations.
}
\label{fig:representation}
\end{figure}

When I say data are a means to an end, I do not mean they are epistemically unimportant; only that they are not themselves the target of inference. Used well, they provide evidence about the process that generated them, and interpreting them therefore requires domain knowledge. A model’s output serves scientific inquiry when, within the interpretive framework of a scientific community, it supports correct inferences about the phenomenon.\footnote{The inferential criterion adopted here has an analogue in \citet{sullivan2023Do}'s interpretative view of representation.}

In the introduction I said that a hallucination is an \textit{error} that is produced by the model, rather than one that is inherited from the training data. An error is a deviation from some standard, and the picture above makes it clear that the target phenomenon, rather than the training data, is the relevant standard. But now we should also ask: what kind of deviations count as errors? That question has no straightforward answer because what counts as an error depends in part on the interpretive practices of the relevant scientific community.

To see why interpretation matters, consider a non-scientific example. The website \texttt{thesecatsdonotexist.com} produces realistic images of cats using a StyleGAN model trained on photographs of real cats. Now suppose you were learning about cats from a sequence of images that included both real photographs and outputs from this model. Under the canonical interpretive scheme for photographs, according to which they depict particular, spatiotemporally located individuals, the StyleGAN images count as misrepresentations. However, if you can reliably identify which images were generated by the StyleGAN model, and qualify any inferences you might be apt to draw from those images in light of that knowledge, they need not cause any epistemic harm. The mere fact that these misrepresentations exist in your data will not prevent you from acquiring new knowledge. After all, the synthetic images still provide accurate information about the statistical properties of cat (or cat-like) images, even though they don't represent any particular animal. The important point is that a model output might count as a misrepresentation according to the standard interpretive scheme for representations of that kind without misleading anyone, as long as we have some method of anticipating the hallucination, and changing our interpretation of the output accordingly.\footnote{This is why there is no contradiction in speaking of detecting, filtering, or qualifying hallucinatory outputs. The propensity to mislead researchers is not part of the definition. In this respect, hallucinations \textit{are} a bit like perceptual illusions. The Müller-Lyer illusion remains an illusion even after one learns that it is illusory. Similarly, a model output that gets recognized as hallucinatory does not automatically lose its status as a hallucination.} 

This puts us in position to distinguish hallucinations from another familiar category of scientific misrepresentation: idealizations. Scientific models include systematic distortions to render phenomena tractable \citep{weisberg2007Three, strevens2016How, sullivan2019Idealizations}. These are \textit{strategic} distortions, designed to facilitate inference. Idealizations can also misrepresent their target, but because they were introduced deliberately, the inferential habits of the researchers are adjusted to take that misrepresentation into account. Hallucinations, by contrast, are \textit{non-strategic} misrepresentations. They are unintentional artifacts of the generative process that demand the construction of an explicit mitigation strategy.

With these distinctions in place, we can now define hallucinations for scientific applications of generative AI. A hallucination is a generative AI model output that satisfies three conditions:
\begin{enumerate}
\item It counts as a misrepresentation of the target system according to the canonical interpretive scheme for outputs of that kind.
\item It is non-strategic: an unintended artifact of the generative process rather than a deliberate idealization.
\item Its misrepresentational status was produced by the model's generative activity, rather than having been inherited from the training data or inference-time inputs.\footnote{Because the definition is restricted to generative AI, the high-dimensional structured character of these outputs is taken as given; see Section~2.1.}
\end{enumerate}

Now that we know what hallucinations are, we are in a better position to understand how they can be managed. The main idea is that we can draw on independent, pre-existing knowledge of the target system to filter or qualify model outputs in ways that prevent possible misrepresentations from misleading us. The next two sections provide case studies that illustrate two rather different ways of doing just that.

\section{Managing hallucination in AlphaFold}

AlphaFold is the most widely known success story about the use of AI in science. The AlphaFold models, built by Google DeepMind, predict folded protein structure from amino acid sequence. AlphaFold 2 achieved unprecedented accuracy on this problem, and the work that led to the model was recognized with one half of the 2024 Nobel Prize in Chemistry, awarded jointly to Demis Hassabis and John Jumper. AlphaFold 3 was built with a larger predictive scope. It covers interactions between proteins and other molecules, including drug-like compounds. This expansion was enabled in part by redesigning the central component that turns the model’s latent representations into candidate three-dimensional molecular structures. The version of that component that appeared in AlphaFold 2 produced a point estimate. The corresponding component in AlphaFold 3, however, is based on a diffusion architecture. Accordingly, it learns a distribution over possible outputs and generates predictions by sampling from that distribution. This makes AlphaFold 3 unambiguously generative in the sense defined in Section 2. Unsurprisingly, therefore, it also introduces the risk of hallucination, as the paper presenting the model explicitly acknowledges:

\begin{quote}
 The use of a generative diffusion approach comes with some technical challenges that we needed to address. The biggest issue is that generative models are prone to hallucination, whereby the model may invent plausible-looking structure even in unstructured regions \citep[p. 496]{abramson2024Accurate}.
\end{quote}

Note that the errors in question do in fact satisfy our three-part definition. First, AlphaFold outputs are conventionally interpreted as representations of a target molecule's three-dimensional structure. A model output that assigns a neat, compact conformation to an intrinsically disordered region is, according to the standard interpretation of chemical-structure representations, a plain misrepresentation of its target. Second, these distortions are non-strategic. They are unintended artifacts of generative sampling, not idealizations introduced to facilitate inference. Finally, these errors are not straightforward reproductions of erroneous training data or inference-time inputs. The diffusion module generates a determinate representation of structure where the available evidence does not support it. In this sense, the misrepresentational content is produced by the generative sampling procedure rather than being inherited from the training data. (This is not meant to imply that the model's disposition to produce hallucinations is causally independent of its training data. All of the model's dispositions are shaped by training.) These errors therefore count as hallucinations in the sense defined above, and consequently give rise to the epistemological puzzle introduced in Section 1, which can be expressed as follows. Given that hallucinations cannot be eliminated wholesale, and that individual hallucinations cannot be traced back to any particular flaw in the model construction process, how can the inferences we make on the basis of such a model be reliable? The answer to this question is that the inferences are made from the larger workflow built around the generative model, and that workflow can be reliable even if the model itself continues to hallucinate. 

To make this clear, let's distinguish between (i) strategies that \textit{prevent} hallucinations (as well as other kinds of error) from being produced in the first place, and (ii) strategies that \textit{flag} defective representations once they occur. Once a defective representation has been flagged, it may be filtered out by an automated procedure before it reaches the user, or, if it is a user-facing output, it might be qualified in some way, so that its defective status is communicated to the user, who can then adjust their interpretation accordingly. Both prevention and flagging operations are crucial to the success of AlphaFold. However, as a response to our epistemological puzzle about reliability, the flagging operations are particularly important, because they show how a model that continues to hallucinate, despite the best efforts of scientists, can nevertheless be embedded in a reliable workflow.  

Now let's examine the problem of hallucination in intrinsically disordered regions of a protein. An intrinsically disordered region is one that has no stable conformation. Most protein segments fold into a particular, predictable shape under normal physiological conditions \citep{nelson2021Lehninger}. An intrinsically disordered region is one that takes on a variety of different shapes depending on subtle variation in the molecular environment, even when that variation remains within the normal range \citep{piovesan2022Intrinsic}. Stable regions tend to pack together tightly when they fold, whereas disordered regions tend to remain relatively extended and exposed to the surrounding molecular environment. This difference is exploited by AlphaFold 3 as a means of distinguishing accurate representations of disordered regions from hallucinations, which typically represent a disordered region as having some particular compact conformation.

With respect to preventing error, one of the key mechanisms in AlphaFold 3 is called \textit{cross-distillation}. The AlphaFold 3 team supplemented the training corpus with structures predicted by another model called AlphaFold-Multimer, which represents unstructured regions as extended loops rather than compact folds. Training on these predictions discourages AlphaFold 3 from inventing determinate structure. The team reports that cross-distillation \enquote{greatly reduced the hallucination behaviour} of the model, but it did not eliminate it \citep[p. 496]{abramson2024Accurate}.   

With respect to flagging error, AlphaFold 3's principal automated mechanism is a process of candidate ranking. AlphaFold 3's diffusion module generates many candidate structures, and this creates a need for a ranking mechanism to determine which candidate is best. That ranking mechanism finds defects in candidate structures, and down-ranks them accordingly.  To do this, the ranking mechanism must combine information from multiple sources. 

One of these sources is called the \textit{confidence head}, a part of the neural network that learns to predict error rather than generate new structure. It is trained to predict discrepancies between model-generated candidates and experimentally derived reference structures. Because its learning objective differs fundamentally from that of the diffusion component, the confidence head acquires sensitivity to errors to which the generative component remains insensitive, and which it therefore continues to produce, despite the best efforts of scientists. 

The ranking mechanism is also informed by theoretical constraints imposed by the model's designers. For example, a steric clash term penalizes candidates containing extensive overlap between atoms in different polymer chains \citep{nelson2021Lehninger, abramson2024Accurate}. Moreover, a solvent-accessibility term gives an advantage to candidates containing more solvent-exposed segments of the protein. The conformations of an intrinsically disordered region tend to be loose, such that more of the protein is exposed to solvents in the environment. This post-hoc advantage given to candidates with more solvent-exposed segments therefore serves to counteract the generative component's disposition to misrepresent disordered regions as compact.     

Candidate ranking improves the reliability of the workflow only if it reliably tracks the accuracy of the candidate structures. The AlphaFold 3 team therefore evaluated the ranking procedure on data independent of that used to train the model. As the number of candidate structures increased, the ranking procedure selected increasingly accurate predictions \citep[Extended Data Fig. 7b]{abramson2024Accurate}. This provides independent evidence that candidate ranking tracks the accuracy of candidate structures, rather than merely reproducing the generative model’s own dispositions.

Once the candidate structures have been ranked, the winner is delivered as the user-facing output. But the workflow reliability is enhanced in another way. Ranking operates on complete structures, and some parts of the selected structure are more reliable than others. To help guide scientists in their interpretation of the output, therefore, the same confidence head also produces a local confidence score called the predicted local distance difference test (pLDDT), which predicts the accuracy of the structural neighborhood around each atom. These scores perform a second kind of flagging operation by allowing users to adjust their interpretation of the selected output \citep{abramson2024Accurate}. 

Like the candidate ranking mechanism, pLDDT scores enhance the reliability of the workflow only if they really track the accuracy of the regions to which they are assigned. The AlphaFold 3 team therefore evaluated pLDDT on proteins that did not appear in the dataset used to train the generative component of the model, testing whether low confidence scores correctly identified disordered regions \citep[Extended Data Fig. 1]{abramson2024Accurate}. This provides independent evidence that low pLDDT scores really do identify intrinsically disordered regions. Scientists can therefore use those scores to distinguish parts of the output that can support inference from those that warrant caution.

Taken together, this family of error-flagging techniques provides the answer to our epistemological puzzle. Scientific inference is not driven directly by the generative component of AlphaFold 3, but by a larger workflow in which different mechanisms are sensitive to different kinds of error. Candidate ranking filters defective structures before they reach the user, while local confidence scores provide an interpretive guide to the varying reliability of different parts of the output structure. The lesson of the AlphaFold 3 story is that reliability does not require hallucination to be eliminated at the level of the generative model (although that is obviously desirable). It can instead be achieved at the level of the workflow, by means of mechanisms that filter and qualify the model’s errors.

\section{From molecules to meteorology}

In this section I will describe a generative method for meteorological forecasting called SEEDS \citep{li2024Generative}, which I have chosen because it presents a useful contrast with AlphaFold. Meteorological forecasting concerns a planet-scale dynamical system that evolves over time and is therefore rather unlike the relatively stable molecular-scale phenomena that AlphaFold is designed to predict. More importantly, the probabilistic form of SEEDS’s output illustrates a way of mitigating the threat of hallucination that is absent from the AlphaFold case.

Some background about weather prediction is in order. Famously, weather systems are chaotic, and this chaos is reflected in the differential equations we use to predict them. Tiny differences in the choice of initial conditions get amplified as the calculation is carried forward in time, and quickly diverge into substantially different alternative future states \citep{lorenz1995Essence}. To deal with the inevitable uncertainty associated with modeling such a chaotic system, weather forecasting typically involves the production of a whole ensemble of such forecasts. Traditional weather prediction models generate an ensemble by computing solutions to the relevant differential equations numerically. The trouble with this method is that a new solution must be computed for each member of the ensemble, which makes the procedure computationally intensive. This becomes particularly problematic when we want to estimate the probabilities of low-frequency events such as cyclones or extreme heat. Estimating probabilities with a useful degree of precision requires increasingly large ensembles as you move out toward the tails of the forecast distribution. There is, therefore, plenty of motivation to design a less computationally intensive alternative.

SEEDS offers such an alternative. It begins with a small number of ensemble members generated by means of a conventional numerical weather prediction model. These seed forecasts are used as inputs that condition the outputs of a trained diffusion model, which has learned subtle statistical regularities in historical data. Then, using the denoising process described in Section 3, the diffusion model generates a plausible version of the global weather at some future date. Each output is a complete, multivariate global weather map at that date. This process is repeated many times, with slight variation in the noise parameter. The result of this process is a new, AI-driven forecast ensemble. 

Does such a model give rise to hallucinations, in our sense? The question is more complicated here than it was in the case of AlphaFold. The canonical interpretation of each member of the ensemble is that it represents one way the weather might be; it is not a deterministic claim about how the future will in fact be. You might think that the fact that an individual forecast makes no deterministic claim about the future means that it cannot count as a hallucination. A hallucination is a misrepresentation, and it is not obvious what would make a single ensemble member count as a misrepresentation. The answer is that each member should correspond to a genuine possibility. It must, in other words, be internally consistent, given the physics of weather. 

Remember that each forecast is a complex object with many variables. The relationships between these variables are bound by various physical constraints. As an example, consider the phenomenon known as hydrostatic balance. Air pressure decreases with altitude, and the pressure gradient created by this decrease generates an upward force on the air. That force is counteracted by gravity, so that the two forces are approximately in equilibrium (which is why the balance is called `static'). This equilibrium means that the drop in pressure seen within a layer is fixed by the weight of the air in that layer. Because warm air is less dense than cold air, it weighs less per unit height. A warm air column must therefore be taller than a cold one to produce the same drop in pressure. A consequence of this is that the temperature of a layer and the height at which a given amount of pressure occurs are nomologically dependent on each other \citep{holton2012Introduction}. When generating forecasts with a diffusion model, however, an individual forecast might assign values to these variables that are individually plausible, but also jointly inconsistent with this nomological dependence constraint. Such a member would represent a physically incoherent state as being a live meteorological possibility, and therefore count as a misrepresentation.

Conditions 2 and 3 of the definition of hallucination are also satisfied. A hydrostatically incoherent ensemble member is a case of \textit{non-strategic} misrepresentation. The properties that would render it incoherent are not deliberately introduced to facilitate some downstream inference. Moreover, if the seed forecasts and relevant training data are coherent, then any incoherent ensemble member will have been created by the diffusion process itself. 

This shows that individual ensemble members \textit{can} be hallucinatory, but doesn't show that any of them actually are. The reason for this is that multivariable incoherence of this sort is subtle and hard to screen for. As I said in Section 1, scientific hallucinations are not typically as conspicuous as an image of a six-fingered hand. An incoherence of the sort at issue here would require specialized diagnostics to test, and there are too many such possible incoherent sets of variables to build a test for each. The authors of the paper acknowledge this in a section called ``Hallucination or in-filling?'' Although they do inspect individual forecasts with a few coherence checks, their reported diagnostics primarily concern evidence for coherence at the ensemble level. The authors consequently acknowledge that the occurrence of hallucinations in individual members cannot be ruled out \citep[p.~7]{li2024Generative}. 

Despite this, scientists can make reliable inferences on the basis of the information furnished by SEEDS. This is made possible in large part by the ensemble format of the output and its probabilistic interpretation. Since any individual hallucination is confined to a particular member of an ensemble, which, in a typical case, has 512 members \citep[p.~2]{li2024Generative}, each member carries only a sliver of the empirical content conveyed by the output. As long as hallucinatory members are not too common and not too correlated, they have little impact on the kind of practical guidance a weather forecast is meant to offer. 

This is a form of what I called ``qualification'' in the previous section. Like the pLDDT scores in AlphaFold, it involves a multipart representational format in which one part qualifies the interpretation of another. The two cases qualify in different ways, however. A pLDDT score is an annotation to the primary output (where ``primary output'' refers to the complete representation minus the pLDDT score itself) that provides an explicit estimate of its reliability. By contrast, couching a representation within an ensemble provides no explicit information about the reliability of the primary output, because the ensemble \textit{is} the primary output. Nothing described so far provides an estimate of the reliability of the ensemble itself. For that, ensemble-level validation techniques are needed. 

Suppose we generate 100 ensemble members and 30 of them say that it will rain. Can we then say that the probability of rain is 30\%? Not on that evidence alone. This is because ensemble members are artifacts of our modeling choices, rather than random samples from a population out there in nature. As \citet[pp.~268--270]{parker2010Predicting} argues, weather probabilities cannot simply be read off from ensemble frequencies. Interpreting ensemble frequencies as probabilities requires additional justification. In the case of SEEDS, the authors post-process their data with bias-correction techniques, and they validate the resulting probabilities against held-out data, using measures that include the continuous ranked probability score (CRPS) and the Brier score \citep[figs.~5--7, pp.~6--7]{li2024Generative}. The mathematical details of these measures are not important here. What matters is that they are part of an epistemic strategy for demonstrating that the ensemble frequencies can be treated as guides to weather probabilities.

Predictive success on historical data provides a reason to expect future success only if there is good reason to think that the conditions represented in the model evaluation are relevantly similar to the future conditions in which the model will be used. In order to know that future conditions are indeed relevantly similar to the conditions in the model evaluation, background knowledge about such conditions is needed. In particular, background knowledge about which conditions actually influence forecasting performance. When that similarity judgment has been made accurately, the historical comparison provides a reason to treat the ensemble frequencies as guides to future weather probabilities. This shows that reliability requires knowledge-informed inference, and this makes it more demanding than brute induction. 

We have now seen two mechanisms that enable SEEDS to support reliable inference despite the threat of hallucination. The ensemble format limits the epistemic significance of any individual hallucination, and the ensemble-level validation supplies positive evidence that the resulting probabilistic forecasts are reliable under the conditions tested. Neither mechanism eliminates hallucination, and unlike the filtering described in the AlphaFold case, neither involves discarding outputs suspected of being inaccurate. Instead, potentially hallucinatory outputs are embedded within a larger inferential structure and assessed at the level at which they support downstream inference. 

\section{Discovery and justification}

I have argued that the threat of hallucination does not undermine the reliability of scientific AI because generative models can be embedded in epistemically robust workflows. Yet one might object that my emphasis on strategies for reliability misses what ought to be the centerpiece of any response to concerns about hallucination: the main epistemic safeguard for scientific AI is post-hoc empirical validation. We trust AlphaFold primarily because we can experimentally test its predictions, and we trust SEEDS because we can wait two weeks and check the temperature.

This objection is inspired by Duede’s [\citeyear{duede2023Deep}] argument that concerns about AI reliability often misunderstand its scientific role. Duede claims that AI is fundamentally a tool for \textit{discovery}, not justification. On this view, AI merely offers heuristic guidance, narrowing the space of hypotheses, which become candidates for belief only through empirical test. The thought echoes Karl Popper’s [\citeyear{popper1959Logic}] influential distinction between the context of discovery and the context of justification: the processes that generate hypotheses may be psychologically or technologically interesting, but epistemology is only concerned with the manner in which they are exposed to empirical testing.

The historical turn in the philosophy of science cast doubt on the sharpness of this division. Discovery and justification are often intertwined. Scientific heuristics are shaped by predictive track record, but also by theoretical expectations, validation practices, and judgments about when outputs are reliable enough to guide further inquiry. AlphaFold was not funded merely because it generated intriguing hypotheses, but because its developers gave theoretically grounded reasons to think, even before large-scale validation, that it could reliably predict biologically plausible structures. 

A related objection is that these strategies have a merely negative cast, since they are concerned with screening error rather than accumulating positive evidence for a model’s output. But, from a reliabilist perspective, that distinction breaks down. Filtering or managing potential errors increases the reliability of an inferential process. And because reliability is what transforms true belief into knowledge, error management bears directly on the epistemic status of model-supported beliefs.

\section{Error mitigation and the construction of reliable science}

Once we conceptualize hallucination as a misrepresentation of the target phenomenon, rather than as a deviation from the model's training distribution, strategies for securing the reliability of scientific inference become visible. These strategies use existing knowledge of the target phenomenon to assess model outputs and determine how those outputs should be filtered or qualified, so that errors in the model output (whether hallucinatory or not) need not propagate into downstream scientific inference. As the cases of AlphaFold and SEEDS both demonstrate, these methods do not eliminate hallucination, but they make it possible to build a reliable workflow, despite the fact that the generative component of that workflow continues to hallucinate.

The rationale behind the techniques described here is partially captured by existing literature on computational reliabilism \citep{duran2018Grounds, duran2025Defense}. In the space that remains, I will briefly explore how the epistemological strategies developed in this paper relate to that literature, with particular focus on the consequences of making the workflow, rather than the model, the central unit of epistemic analysis. 

Computational reliabilism says that computational systems can be judged to be reliable, despite the fact that they are epistemically opaque, by relying on what \citet{duran2026Transparency} calls \textit{reliability indicators}. These include benchmark tests, calibration metrics, and many other kinds of scientific practice. The category of reliability indicators is so broad, in fact, that even the epistemic standards of the scientific community are included. \citet{alvarado2026Challenges} puts pressure on Dur\'{a}n's account, arguing that he has underestimated the difficulty of the problems caused by epistemic opacity. Alvarado lists three such problems, of which the third is most relevant here. It is that, in order to make a judgment of reliability strong enough to use the model in epistemically rigorous contexts, you need to have knowledge about how and why the model fails. In deep neural networks, Alvarado argues, this kind of knowledge is not generally possible, because errors cannot generally be traced to any particular step in the computation that produced them.   

Alvarado's analysis dovetails with the analysis of hallucination given here. Hallucinations cannot be traced to any localized defect in the computational process that produced them. The way Alvarado sees it, this leaves us with a dilemma. Either computational reliabilism has to incorporate endogenous information about the nature and source of model errors, and thereby confront opacity head-on, or it has to admit that the kind of reliability assessments described by Dur\'{a}n will remain inadequate for rigorous science.   

The case studies here suggest things are not quite that bad. A third possibility is made available when we reject the assumption that the model itself is the unit at which reliability must be assessed. Instead, we set our sights on constructing a workflow in which model errors are already taken into account.\footnote{The word ``model'' has many meanings. One might worry that I am trading on a particular parochial analysis of the term. I'm not. I'm trading instead on the claim defended in Section 2 that generative AI models will inevitably hallucinate. The relevant contrast is between the generative model (or model component, if you prefer), and the workflow in which it is embedded.} This does not require that errors are traced back to their source, as the first horn of the dilemma seems to suggest. It only requires that we have enough information about the conditions in which they occur, and enough independent knowledge of the target system, to filter and qualify the model outputs appropriately. The workflow is the relevant unit in this process because it is the site at which the filtering and qualifying interventions are made. In recent decades, philosophers of science have come to recognize models and simulations as valid units of epistemic analysis, alongside theories \citep{morgan1999Models, humphreys2009Philosophical}. The case studies here suggest that workflows ought to be added to this list. 

The workflow-centric strategies presented here also shed new light on the relation between reliability indicators and reliability itself. Standard process reliabilism says that a belief can be warranted by a reliable process even if the believer is not in a position to certify that the process is reliable \citep{goldman1979What}. This logical independence does not prevent knowledge of reliability from playing a causal role in the construction of a reliable process. At an initial stage, evidence about the performance of the generative model, interpreted against a background of pre-existing knowledge of the target system, allows researchers to determine which uses of its outputs can be made safely, and under what conditions. Researchers then use that knowledge to qualify and filter the outputs of the generative model. These qualifying and filtering procedures are themselves parts of the workflow, which can therefore be more reliable than the model considered in isolation. On this view, the reliability of a process and knowledge of its reliability remain logically distinct. This means that a user can acquire knowledge through the workflow without knowing that the workflow is reliable. There is therefore no higher-order knowledge condition imposed, which is appropriate, since those higher-order conditions on knowledge, of which the KK principle is the most well-known, violate the externalist spirit of the reliabilist view. Nevertheless, knowledge of reliability is causally, rather than logically, implicated in the improvement of the workflow over time.

A final point is that the reliability-enhancing strategies described here require established scientific knowledge of the target system. We saw this in AlphaFold's use of knowledge about intrinsically disordered regions to govern the interpretation of pLDDT scores, and in SEEDS's use of knowledge about atmospheric dynamics and ensemble forecasting to qualify individual forecasts. Where comparable knowledge is lacking, it may be impossible to construct a reliable workflow around a generative AI model. This dependence reflects the Deweyan insight expressed in the epigraph: reliable inference depends on converting mature knowledge into procedures for filtering and qualifying model outputs. Generative AI makes that work harder, but the case studies show that, under the right conditions, scientists can still find ways of \textit{making sure}.

\clearpage
\bibliography{Hallucination}
\end{document}